\newif\ifjpsj
\jpsjfalse

\ifjpsj
 \documentclass[twocolumn,letter]{jpsj3}
\else
 \documentclass[twocolumn,amsmath,amssymb,longbibliography,prb]{revtex4-2}
 \usepackage[dvipdfmx]{graphicx}
 \usepackage{hyperref, xcolor}
 \hypersetup{
 	colorlinks=true,
 	citecolor={magenta},
 	urlcolor={blue!70!black}
 }
\fi

\newif\iftitle
\titletrue

\usepackage{bm}
\usepackage{color}
\usepackage{braket}
\usepackage{lipsum}
\usepackage{mathdots}

\usepackage{xspace}
\usepackage{ulem}
\usepackage{url}

\newcommand  {\eqn}[1]{(\ref{eqn:#1})} 
\renewcommand{\(}     {\left(}
\renewcommand{\)}     {\right)}

\renewcommand{\_}[1]  {_\textrm{#1}}

\begin{document}

\title{
Backsolution: A Framework for Solving Inverse Problems via Automatic Differentiation
}

\newcommand{\authA}{Koji Kobayashi\thanks{k-koji@sophia.ac.jp}}
\newcommand{\authB}{Tomi Ohtsuki\thanks{ohtsuki@sophia.ac.jp}}

\newcommand{\affiA}{Physics Division, Sophia University, Tokyo 102-8554, Japan}

\ifjpsj 
 \author{\authA and \authB}
 \inst{\affiA
      }
\else
 \author{\authA}
 \author{\authB}
 \affiliation{\affiA}
\fi

\newcommand{\abstbody}{
 We present a simple yet powerful framework for solving inverse problems by leveraging automatic differentiation.
 Our method is broadly applicable whenever a smooth cost function can be defined near the true solution, and a numerical simulator is available.
 As a concrete example, we demonstrate that our method can accurately reconstruct the spatial profiles 
 (potential or magnetization landscapes)
in a conductor from magnetotransport measurements.
 Even if the given data are insufficient to uniquely determine the profiles, the same framework enables effective reverse modeling.
 This method is general, flexible, and readily adaptable to a broad class of inverse problems across condensed matter physics and beyond.
}
\ifjpsj
 \abst{\abstbody}
\else
 \begin{abstract}
  \abstbody
 \end{abstract}
\fi

\maketitle

%%%%%%%%%%%%%%%%%%%
%%%%%  INTRO  %%%%%
%%%%%%%%%%%%%%%%%%%
\iftitle
 \section{Introduction}
\else
 \textit{Introduction.}
\fi
%%% generic introduction to inverse problems %%%
 Inverse problems play a central role in many areas of physics, enabling us to infer hidden internal properties of materials from observed data.
 Except for well-posed linear cases, they are rarely solvable analytically and require advanced numerical techniques: such as iterative solvers, Markov chain Monte Carlo, regularization, Bayesian inference, neural networks, and hybrid methods \cite{Linden95maximum, Mayer04bayesian, Toussaint11bayesian, Vocciante15ageneral, Tamura17method, Raissi19physics, Kades20spectral, Webb25large}.
 Despite decades of research, many inverse problems remain practically challenging.

%%% Determining potential %%%
 Reconstructing invisible internal profiles of materials or devices is one of those difficult classes of problems.
 In traditional approaches, only averaged impurity profiles (e.g., density or intensity) were estimated from energy-dependent conductance \cite{Mukim20disorder}.
 More recently, machine learning techniques have enabled the prediction of more detailed features, such as the positions of antidot potentials from magnetofingerprints \cite{Daimon22deciphering,Yokoi25} or potential landscapes from scanning gate microscopy images \cite{Percebois23reconstructing}.
 However, these approaches are highly specialized for the specific problems and have limited generalizability, leaving many inverse problems still out of reach.

%%% inverse design %%%
 A related concept is inverse design, which aims to optimize structures or model parameters to achieve desired outcomes \cite{Franceschetti99,
 Fujita18, Yu21learning, Inui23inverse, Williams23automatic, Inui24inverse, Hirasaki24inverse}.
%%% automatic differentiation %%%
 Recent progress in this area has been driven by advances in deep learning, particularly through optimization techniques like backpropagation \cite{Amari67, Rumelhart1986},
which relies on automatic differentiation (AD).
 The AD has been applied to inverse designing of effective Hamiltonian parameters (e.g., spin-spin or spin-orbit couplings) \cite{Yu21learning, Inui23inverse, Williams23automatic, Inui24inverse} and of potential landscape \cite{Hirasaki24inverse}.
 This has extended the reach of gradient-based optimization beyond manually or perturbatively differentiable problems \cite{Fujita18}.
 Nevertheless, the AD-based method can be less efficient than Bayesian methods \cite{Yu21learning}, especially for a few parameter optimization, and its impact on inverse design in physics remains limited.

%%% Outline %%%
\iftitle
 In this paper, 
\else
 In this Letter,
\fi
we introduce a simple and versatile method for solving inverse problems using AD, which we call the ``backsolution'' method, after the backpropagation technique.
 The method is scalable and effective even for problems involving over $100$ unknown parameters, where conventional Bayesian methods become computationally inefficient.
 We demonstrate that the backsolution method can accurately reconstruct potential and magnetization profiles from magnetotransport data.
 Furthermore, we apply it to an ill-posed problem: inferring candidate lattice structures from Fourier images of amorphous-like systems.
 While the examples in this Letter are minimal, our method is broadly applicable and can be integrated with existing optimization algorithms.

%%%%%%%%%%%%%%%%%%%%
%%%%%  METHOD  %%%%%
%%%%%%%%%%%%%%%%%%%%
\iftitle
 \section{Method}
\else
 \textit{Method.}
\fi
\iftitle
 \subsection{Backsolution}
\else
% \textit{Backsolution---}%
\fi
 We begin by formally introducing the backsolution method, which solves inverse problems as optimization tasks of parameters $\bm{p} \equiv \{p_1,p_2,...,p_n\}$.
 To infer the true hidden parameters $\bm{p}^\mathrm{true}$ from an observation dataset $\{\bm{x}_i, y^{\mathrm{true}}_i\}$ $(i=1,...,N\_{data})$,
we minimize a suitable cost function $f(\bm{p})$.
 (For simplicity, we assume that each $y^{\mathrm{true}}_i$ is a scalar.)
 We employ the mean squared error as the cost function,
%___ eqn:MSE ___%
\begin{align}
 f(\bm{p}) &= \frac{1}{N\_{data}} \sum_{i=1}^{N\_{data}} [g(\bm{x}_i; \bm{p}) - y^{\mathrm{true}}_i]^2,
 \label{eqn:MSE}
\end{align}
%---%
where $g$ is a simulator that returns the observable $y_i$ for a given set of variables $\bm{x}_i$ 
and the parameters $\bm{p}$.
 By providing an initial guess for $\bm{p}$,
we compute the gradient vector of the cost function,
$\nabla f(\bm{p})=\(
   \frac{\partial f(\bm{p})}{\partial p_1},
   \frac{\partial f(\bm{p})}{\partial p_2},
   ...,
   \frac{\partial f(\bm{p})}{\partial p_n} \)$,
using AD.

%% JAX %%
 In this work, we use the \texttt{grad} function from the JAX library \cite{jax2018github} to obtain the gradient,
%___ eqn:autograd ___%
\begin{align}
 \(\frac{\partial f(\bm{p})}{\partial p_1},
   \frac{\partial f(\bm{p})}{\partial p_2}, ...,
   \frac{\partial f(\bm{p})}{\partial p_n} \)
 = \mathtt{jax.grad}(f)(\bm{p}).
 \label{eqn:autograd}
\end{align}
%---%
 The \texttt{grad} function efficiently and precisely computes all $n$ partial derivatives in a single forward calculation of $f(\bm{p})$.
 This provides scalability for the backsolution method and is a significant advantage over numerical differentiation (finite difference), which typically requires $2n$ forward calculations of $f$.
 However, using AD with JAX introduces certain constraints on the coding.
 The cost function including the simulator must be written in a JAX-compatible manner, which can limit flexibility.
 For instance, \texttt{for} and \texttt{if} statements that dynamically control the flow are not easily supported.
 Despite these limitations, JAX offers high compatibility with NumPy functions and supports GPU acceleration, making it a practical choice for large-scale computation.

 Once the gradient vector is obtained,
the parameters $\bm{p}$ can be updated to minimize the cost function.
 For the update algorithm, we employ simple gradient descent:
%___ eqn:descent ___%
\begin{align}
 \bm{p}^\mathrm{new} = \bm{p}^\mathrm{old} - r \nabla f({\bm p^\mathrm{old}}),
 \label{eqn:descent}
\end{align}
%---%
where $r$ is an appropriate learning rate.
 Although advanced optimization algorithms may improve computational efficiency,
we find that the backsolution method successfully works even with a simple implementation.

\begin{figure}[tbp]
 \centering
  \includegraphics[width=1\linewidth]{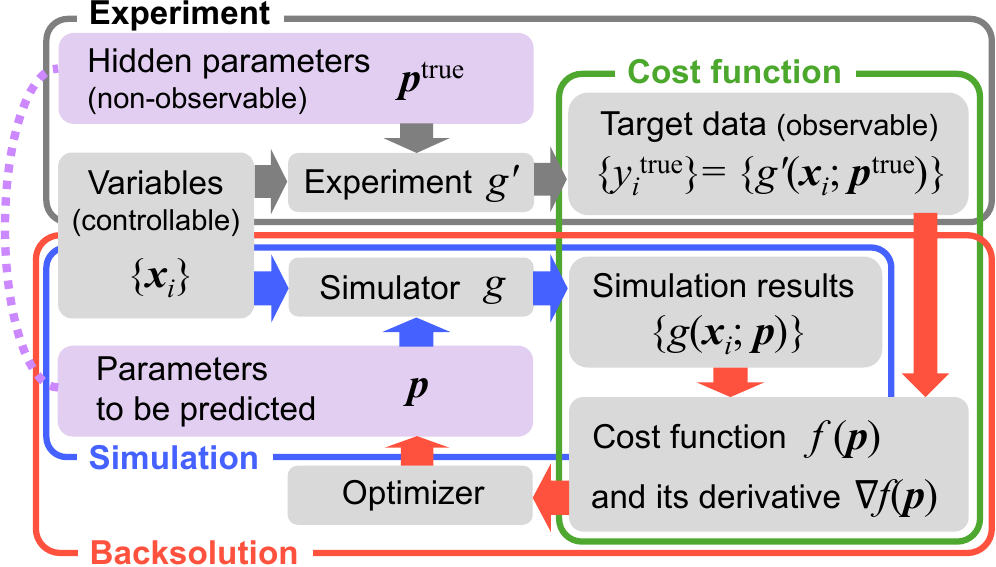}
\caption{
  Schematic flow of the backsolution.
  The flow with gray arrows prepares target data $\{y^\mathrm{true}_i\}$.
  The flow with blue arrows represents the forward calculation: simulation and evaluation of cost function.
  The flow with red arrows updates parameters $\bm{p}$ using the derivative of cost function $\nabla f(\bm{p})$ obtained via AD.
  The forward calculation and parameter update are repeated until convergence.
}
\label{fig:schematic}
\end{figure}

%%% Flow of Backsolution %%%
 Next, we describe the backsolution framework step by step (see Fig.~\ref{fig:schematic}).
 (1) Preparation of target dataset 
$\{\bm{x}_i,y^\mathrm{true}_i\}$%
---%
 Prepare target data 
 $\{y^\mathrm{true}_i\}$ 
 through experiment, $y^\mathrm{true}_i = g'(\bm{x}_i; \bm{p}^\mathrm{true})$.
 The variables $\{\bm{x}_i\}$ should be chosen in a wide range and with a sufficient density for efficient convergence.
 In this work, we numerically generate the data
 for demonstration purposes.
 (2) Initialization of parameters---%
 Specify the hidden parameters $\bm{p}^\mathrm{true}$ to be predicted and prepare an initial guess of $\bm{p}$.
 While random initialization is available, a well-informed guess can significantly accelerate convergence.
 (3) Simulation and cost evaluation---%
 Run a forward calculation of the simulator $\{g(\bm{x}_i;\bm{p})\}$ using the current parameters.
 Then evaluate the cost function $f(\bm{p})$ along with its gradient $\nabla f(\bm{p})$ through AD.
 (4) Parameter update---%
 Update the parameters $\bm{p}$ to reduce the cost $f$.
 Accept the update if the cost decreases.
 Otherwise, retry with a smaller learning rate $r$.
 (5) Iteration---%
 Repeat steps (3) and (4) until the cost converges.
 If the cost stagnates at a relatively large value, restart the process from step (2) with a different initial guess.

%===== FIG =====%
%===== FIG =====%

%%% TMM %%%
\iftitle
 \subsection{Transport simulator: Transfer matrix method}
\else
 \textit{Transport simulator.}
\fi
 In the following examples, we use the transfer matrix method \cite{Pendry92universality, Slevin01} as the transport simulator, which is the fastest algorithm to compute two-terminal conductance.
 The transfer matrix method yields the transmission coefficient matrix $\bm{t}$,
from which the two-terminal conductance $G$ (in units of $\frac{e^2}{h}$) is calculated using the Landauer formula,
%___ eqn:Landauer ___%
\begin{align}
 G &=  \mathrm{Tr}\,{\bm{t}^\dag \bm{t}}.
 \label{eqn:Landauer}
\end{align}
%---%

 To describe the transfer matrix method,
we define the right transfer vector $\Psi$ and the left transfer vector $\Phi^\dag$,
%___ eqn:tv ___%
\begin{align}
 \Psi_x = \begin{pmatrix}
  \psi_x \\
  M_{x-1}^\dag \psi_{x-1}
 \end{pmatrix}, \quad
 \Phi_x^\dag = \begin{pmatrix}
  -\psi_{x-1}^\dag M_{x-1} &  \psi_{x}^\dag
 \end{pmatrix},
 \label{eqn:tv}
\end{align}
%---%
where $\psi_x$ is the wavefunctions at a slice $x$, and
$M_{x} = \braket{x|H|x+1}$ is the hopping matrix between adjacent slices.
 A layer transfer matrix relates the transfer vectors between adjacent slices,
such that $\Psi_{x+1} = T_{x} \Psi_{x}$
and $\Phi_{x}^\dag = \Phi_{x+1}^\dag T_{x}$,
and thus it is written as,
%___ eqn:Tx ___%
\begin{align}
 T_x  &=  
 \begin{pmatrix}
  -M_{x}^{-1} \( \braket{x|H|x} - E \) & -M_{x}^{-1} \\
  M_{x}^\dag & \bm{0}
 \end{pmatrix},
 \label{eqn:Tx}
\end{align}
%---%
where $E$ is the Fermi energy.
 The transfer matrix $T$ for a system of length $L_x$ is the product of the layer transfer matrices,
%___ eqn:T ___%
\begin{align}
 T &=  T_{L_x} T_{L_x-1} \cdots T_1.
 \label{eqn:T}
\end{align}
%---%

%%% lead for TMM %%%
 From the correspondence between the transfer matrix and the scattering matrix,
we understand that the transmission coefficient matrix $\bm{t}$ in Eq.~\eqn{Landauer} is given as,
%___ eqn:transinv ___%
\begin{align}
 \bm{t} &= 
  (U_{\mathrm{L,out}}^\dag \ T \ V_{\mathrm{R,in}})^{-1} \ ,
 \label{eqn:transinv}
\end{align}
%---%
where $U_{\mathrm{L,out}}^\dag$ and $V_{\mathrm{R,in}}$ are the left and right transfer vectors corresponding to the outgoing and incoming eigenmodes in the left and right leads, respectively.
 We can distinguish the right- and left-going (or evanescent) modes using the fact that the current $J_x$ is written in terms of the transfer vectors as,
%___ eqn:current ___%
\begin{align}
 J_x 
 &=  \Psi_{x}^\dag  
 \begin{pmatrix}
     \bm{0} & -i I \\
  i I &    \bm{0} \\
 \end{pmatrix}
 \Psi_{x}
 =  \Phi_{x}^\dag  
 \begin{pmatrix}
     \bm{0} & -i I \\
  i I &    \bm{0} \\
 \end{pmatrix}
 \Phi_{x},
 \label{eqn:current}
\end{align}
%---%
where $I$ is an %$L_y\times L_y$
identity matrix.

%%% perfect lead %%
 To compute the conductance, we have to specify the propagating modes in leads $U_{\mathrm{L,out}}^\dag$ and $V_{\mathrm{R,in}}$.
 As a simple realization of the leads, we use the perfectly metallic parallel one-dimensional wires
\cite{Kramer05random, Kobayashi13disordered} with hopping $-t\_{lead}$,
so that the transfer vectors in the leads are 
%___ eqn:plead ___%
\begin{align}
 V(t\_{lead},k) &= \left|2 t\_{lead} \sin{k}\right|^{-1/2}
  \begin{pmatrix}
      I  \\
   -t\_{lead} e^{-ik} I
  \end{pmatrix}, \\
 U^\dag(t\_{lead},k) &= i\left|2 t\_{lead} \sin{k}\right|^{1/2}
  \begin{pmatrix}
   t\_{lead} e^{+ik} I  \,&\,  I
  \end{pmatrix},
 \label{eqn:plead}
\end{align}
%---%
where $k$ is the wavenumber of the mode.
 Modes with negative $t\_{lead} \sin k$ correspond to left-going modes and are used for $U^\dag_{\mathrm{L,out}}$ and $V_{\mathrm{R,in}}$.
 We set $t\_{lead}=1$.

 In practice, for numerical stability, the transfer matrix method requires orthonormalization of transfer vectors
after every $S\_{D}\lesssim 10$ layers (we use $S\_{D} = 6$ in this work).
 This is typically done via QR decomposition, so that Eq.~\eqn{T} is written as,
%___ eqn:QR ___%
\begin{align}
 &\bm{t} = R_1^{-1} \cdots R_{N\_{D}}^{-1}
           ( V_{\mathrm{L,out}}^\dag \ 
            T_{L_x} \cdots T_{N\_{D}S\_{D}+1}\ Q_{N\_{D}})^{-1}, \\
 &Q_j R_j = 
 \begin{cases}
   T_{S\_{D}} \cdots T_1 U_{\mathrm{R,in}} 
   & (j=1), \\
   T_{j S\_{D}} \cdots T_{(j-1)S\_{D}+1} Q_{j-1}
   & (j>1),
 \end{cases}
 \label{eqn:QR}
\end{align}
%---%
where $Q_j$ is orthonormal and $R_j$ is upper-triangular matrices.
 The number of decompositions $N\_D$ is the largest integer such that $N\_D S\_D+1$ does not exceed $L_x$.
 We use the function \texttt{jax.numpy.qr} in the JAX library for the QR decomposition compatible with AD.

%%%%%%%%%%%%%%%%%%%%
%%%%%  RESULT  %%%%%
%%%%%%%%%%%%%%%%%%%%
\iftitle
 \section{Backsolution for spatial profiles}
\else
% \textit{Results---}%
\fi

%%% Pote %%%
\iftitle
 \subsection{Prediction of potential profile}
\else
 \textit{Prediction of potential profile.}
\fi
 As a first example, we demonstrate that the backsolution method can accurately reconstruct the potential profile of a conductor from magnetotransport data.
 We consider a two-dimensional tight-binding model of $L_x \times L_y$ sites with nearest-neighbor hopping and on-site potential $v_{x,y}$,
%___ eqn:Hpote ___%
\begin{align}
 H &= \sum_{x=1}^{L_x-1} \sum_{y=1}^{L_y} \(\ket{x+1,y} \bra{x,y} + \mathrm{h.c.} \)  \nonumber\\
   &\ + \sum_{x=1}^{L_x} \sum_{y=1}^{L_y-1} \!\! \(\ket{x,y+1} e^{2\pi i\phi x} \bra{x,y} + \mathrm{h.c.} \) \nonumber\\
   &\ + \sum_{x=1}^{L_x}\, \sum_{y=1}^{L_y} \ket{x,y} v_{x,y} \bra{x,y},
 \label{eqn:Hpote}
\end{align}
%---%
where $\phi$ is the magnetic flux in units of flux quantum $h/e$.
 The nearest-neighbor hopping energy is taken as the energy unit.
 In the following, we set $L_x = L_y = 12$.

\begin{figure}[tbp]
 \centering
  \includegraphics[width=0.95\linewidth]{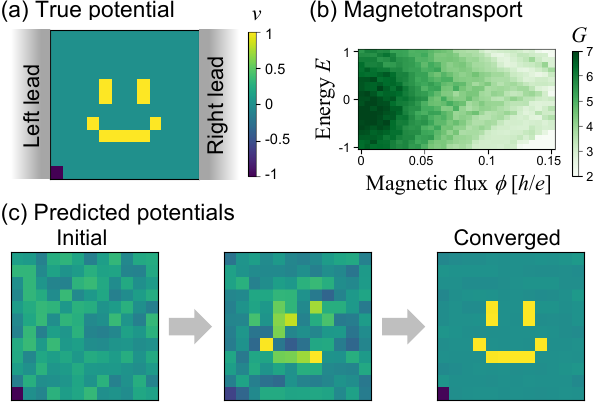}
\caption{
  (a) True potential structure $\bm{v}^\mathrm{true}$, corresponding to the hidden parameter $\bm{p}^\mathrm{true}$ to be predicted.
  (b) Heat map of magnetotransport $G(\phi, E;\bm{v}^\mathrm{true})$ in the $\phi$-$E$ plane, used as the target data 
  $\{y^\mathrm{true}_i\}$.
  (c) Evolution of the predicted potential structure from left, the initial guess, to right, the converged result with final cost $f\simeq 10^{-6}$.
}
\label{fig:pote}
\end{figure}

 The true potential structure $\bm{v}^\mathrm{true}\equiv \{v_{x,y}^\mathrm{true}\}$ is shown in Fig.~\ref{fig:pote}(a).
 Note that $\bm{v}^\mathrm{true}$ corresponds to the hidden parameter $\bm{p}^\mathrm{true}$ in Fig.~\ref{fig:schematic}
and is assumed to be unknown during the backsolution process.
 We choose the variables $\{\bm{x}_i\}=\{\phi,E\}$, where magnetic flux $\phi=[0,0.15]$ with 31 mesh points and Fermi energy $E=[-1,1]$ with 21 mesh points ($N\_{data} = 31 \times 21 = 651$).
 As the target data $\{y^\mathrm{true}_i\}$,
we use the magnetotransport $G(\phi, E; \bm{v}^\mathrm{true})$,
as shown in Fig.~\ref{fig:pote}(b).
 We fix the lead parameter $k=\pi/2$, although it could also be treated as hidden parameters.

 We take the potential $\bm{v}$ as the parameters to be predicted $\bm{p}$.
 The total number of the parameters is $n = L_x\times L_y = 144$.
 The initial guess of $\bm{v}$ is shown in the left panel of Fig.~\ref{fig:pote}(c), consisting of randomly distributed numbers with a fixed `notch' $v_{0,0} = -1$ (introduced for numerical stability, as discussed later).
 Using the transport simulator, we compute $G(\phi, E; \bm{v})$, evaluate the cost function $f(\bm{v})$ as defined in Eq.~\eqn{MSE}, and obtain its gradient via AD.
 Then, we iteratively update $\bm{v}$ to minimize $f(\bm{v})$.
 After some iterations ($\sim 10^3$ epochs in this example),
the optimization converges to a small value (we truncated at $f < 10^{-6}$) at a specific potential structure, as shown in the right panel of Fig.~\ref{fig:pote}(c).
 Remarkably, the backsolution method successfully reconstructs the true potential profile using only two-terminal transport data without using microscopes.
 We have also confirmed that the method is robust in the presence of small noises $\delta G < 0.1 [e^2/h]$ in the target data.

%===== FIG =====%
%===== FIG =====%

%%% DW %%%
\iftitle
 \subsection{Prediction of magnetization profile}
\else
 \textit{Prediction of magnetization profile.}
\fi
 The backsolution method can predict any property encoded in the target data.
 As a second example, we demonstrate the reconstruction of a magnetization profile from magnetotransport measurements.
%% model %%
 We consider a tight-binding model with spin degrees of freedom and exchange coupling $J$ to the magnetization $\bm{m}_{x,y}=(\sin\theta_{x,y},0,\cos\theta_{x,y})$,
%___ eqn:Hpote ___%
\begin{align}
 \!H &= \sum_{x=1}^{L_x-1} \sum_{y=1}^{L_y} \(\ket{x+1,y} \sigma_0 \bra{x,y} + \mathrm{h.c.} \)  \nonumber\\
   &+ \sum_{x=1}^{L_x} \sum_{y=1}^{L_y-1} \(\ket{x,y+1} e^{2\pi i\phi x} \sigma_0 \bra{x,y} + \mathrm{h.c.} \) \nonumber\\
   &+ \sum_{x=1}^{L_x} \sum_{y=1}^{L_y} \ket{x,y} J\( \sigma_z \cos\theta_{x,y} +  \sigma_x \sin \theta_{x,y} \)\bra{x,y},
 \label{eqn:Hmag}
\end{align}
%---%
where $\sigma_\mu$ are Pauli matrices ($\mu=x,z$) and $\sigma_0$ is the $2\times2$ identity matrix.
 We set the exchange coupling $J=0.5$.

\begin{figure}[tbp]
 \centering
  \includegraphics[width=0.95\linewidth]{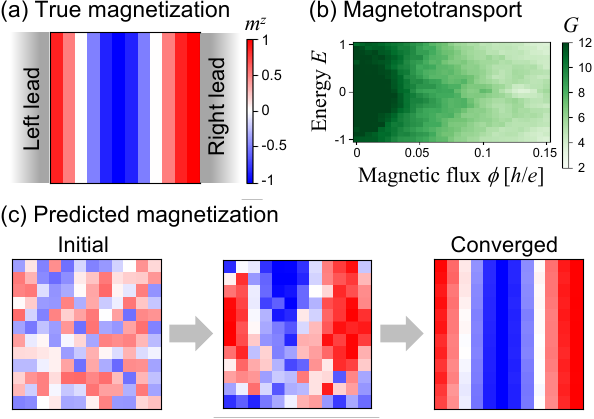}
\caption{
  (a) True magnetization structure
  $m^z_{x,y} = \cos{\theta^\mathrm{true}_{x,y}}$.
  The magnetization angle $\bm{\theta}^\mathrm{true}$ corresponds to $\bm{p}^\mathrm{true}$.
  (b) Heat map of magnetotransport $G(\phi,E;\bm{\theta}^{\mathrm{true}})$ in the $\phi$-$E$ plane, used as the target data $\{y^\mathrm{true}_i\}$.
  (c) Evolution of the predicted magnetization structure from left, the initial guess, to right, the converged result.
}
\label{fig:mag}
\end{figure}

 The true magnetization structure $\bm{m}^\mathrm{true}_{x,y}$ is shown in Fig.~\ref{fig:mag}(a).
 The corresponding target data $G(\phi,E; \bm{\theta}^\mathrm{true})$ with $N\_{data}=651$ are prepared using the transfer matrix method [see Fig.~\ref{fig:mag}(b)].
 To make the conductance asymmetric for $+m_z$ and $-m_z$, we set the lead parameters $k_\uparrow=0.5$ and $k_\downarrow=0.25$, which are the wavenumbers of up- and down-spin modes, respectively.

%% result %%
 Taking $\{\theta_{x,y}\}$ as the parameters to be predicted $\bm{p}$ with $n=144$,
 we apply the backsolution method in the same manner as in the previous example.
 Figure~\ref{fig:mag}(c) shows the evolution of the predicted magnetization profile,
 which successfully reproduces the true magnetization.

%===== FIG =====%
%===== FIG =====%

%%%%%%%%%%%%%%%%%%%%%%%%
%%%%%  DISCUSSION  %%%%%
%%%%%%%%%%%%%%%%%%%%%%%%
\iftitle
 \section{Discussions} \label{sec:discussion}
 \subsection{Limitations}
\else
 \textit{Limitations.}
\fi
 Let us discuss the limitations and practical considerations of the backsolution method.
 One common issue is convergence to a local minimum, which can lead to incorrect predictions.
 Fortunately, this problem is easy to detect early in the process by monitoring the cost function. 
 Since the cost function [Eq.~\eqn{MSE}] directly reflects the accuracy of the solution,
a tendency to converge to a relatively large value typically indicates that the initial guess was suboptimal.
 To address this, a Monte Carlo approach can be used to explore alternative initial states.
 Further techniques for improving convergence are established in the broader context of optimization, and we do not go into detail here.
 We emphasize, however, that all results presented in this Letter were obtained within minutes on a laptop.

%%
%\subsubsection{Ambiguity}
 Another limitation is the non-uniqueness of the solution when the cost function has multiple global minima.
 Typically, this occurs when the target data are invariant under some symmetry about hidden parameters.
 For example, in the case shown in Fig.~\ref{fig:pote},
 the two-terminal conductance remains unchanged under mirror reflections of the potential profile about the $x$ and $y$ axes.
 To avoid such ambiguities in the predicted potential, we introduced a notch in both the sample and the initial guesses.

%%% COST %%%
\iftitle
 \subsection{Applicability}
\else
 \textit{Applicability.}
\fi
 We discuss the conditions under which the backsolution method is applicable.
 For stable convergence, the cost function must be smooth in the vicinity of the global minimum.
 (Importantly, this does not require differentiability across the entire parameter space.)

\begin{figure}[tbp]
\centering
 \includegraphics[width=1\linewidth]{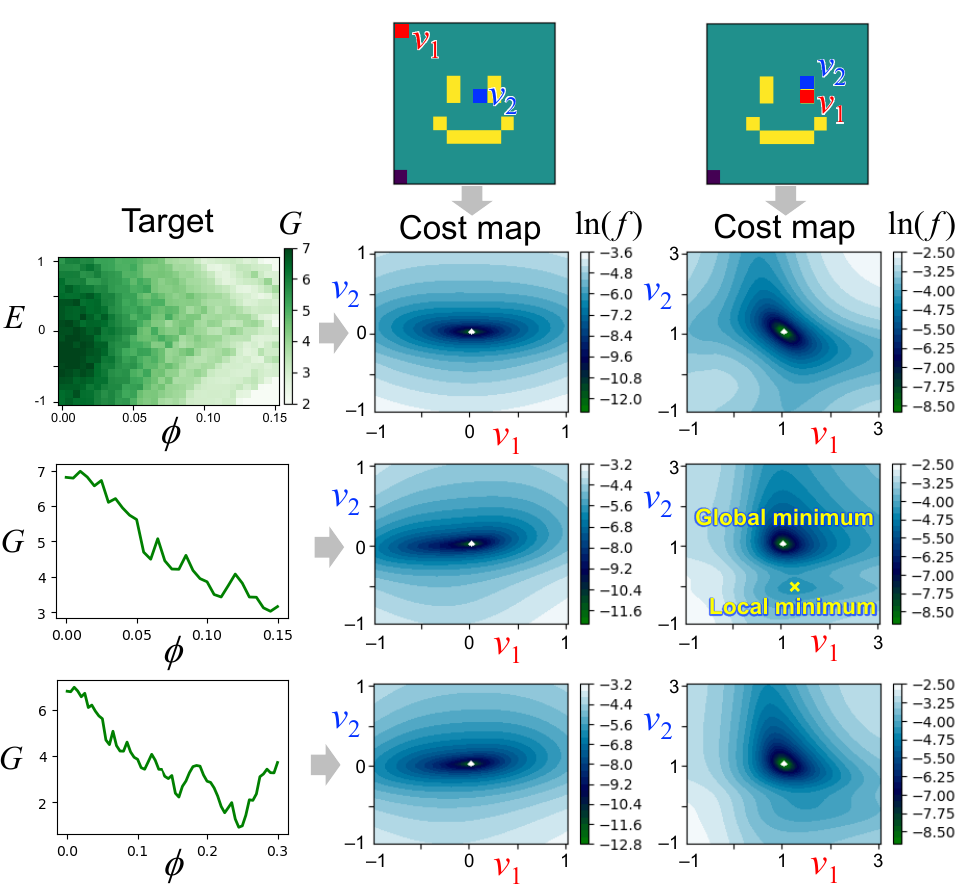}
\caption{
 Left column: target magnetotransport data.
 Center and right columns: cost functions evaluated for $v_1$ and $v_2$ (potential values of two selected sites indicated in the top row).
 Second row: example from Fig.~\ref{fig:pote}, showing a smooth cost function.
 Third row: insufficient target data (only $E=0$ with sparse $\phi$) leads to a local minimum near the global one.
 Bottom row: increasing the mesh points and range of $\phi$ restores the monotonicity of the cost function.
}
\label{fig:costmap}
\end{figure}

 Whether this condition is satisfied is not always obvious from the problem setup alone.
 In our example, neither the potential profiles nor the target data are smooth due to universal conductance fluctuations \cite{Lee85, Lee87}, as shown in the left column of Fig.~\ref{fig:costmap}.
 Nevertheless, the corresponding cost function is smooth as illustrated in the center and right columns of Fig.~\ref{fig:costmap},
 where the cost is plotted as a function of two selected parameters, $v_1$ and $v_2$.
 To improve the smoothness and monotonicity of the cost function towards the global minimum,
we used a conductance map in the $\phi$-$E$ space as the target data.
 When the data are insufficient, the cost function may exhibit local minima near the global one.
 For example, using a sparse mesh in $\phi$ at fixed energy $E=0$, 
a local minimum appears around $(v_1, v_2)\simeq (1.2,0.0)$, while the true minimum is at $(1,1)$, as shown in the right column in the third row of Fig.~\ref{fig:costmap}.
 However, even when the energy is fixed at $E=0$, a sufficiently dense mesh over a wide range of $\phi$ can restore the smoothness and monotonicity of the cost function, as shown in the bottom row of Fig.~\ref{fig:costmap}.
 In such cases, the backsolution method works well under a suitable initial guess.

%===== FIG =====%
%===== FIG =====%

%%%%%%%%%%%%%%%%%%%%%%%%%%%
%%%%% Inverse Fourier %%%%%
%%%%%%%%%%%%%%%%%%%%%%%%%%%
\iftitle
 \section{Application to reverse modeling}
\else
 \textit{Prediction of atomic positions}.
\fi
 As a third example,
we demonstrate that even when the available data are insufficient to uniquely determine the hidden parameters,
the backsolution method can still generate plausible candidates that are consistent with the observations.
 In this sense, the method also serves as a powerful tool for reverse modeling.

 We consider the inverse problem of reconstructing a disordered lattice structure from its momentum-space image.
 Although such an image does not contain enough information to uniquely recover the true atomic positions, we can generate lattice configurations that exhibit similar structural characteristics.

\begin{figure}[tbp]
 \centering
  \includegraphics[width=1\linewidth]{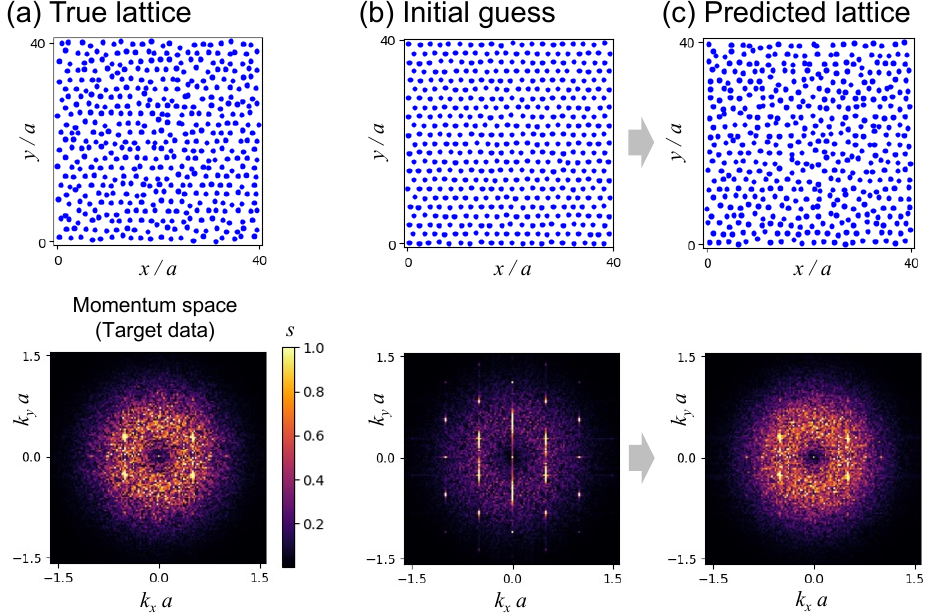}
\caption{
  (a) True, (b) initial, and (c) predicted lattice structures.
  Top panels are atomic positions in real space.
  Bottom panels are momentum-space images $\{s_{k_x,k_y}\}$
  [see Eq.~\eqn{s}].
}
\label{fig:Fourier}
\end{figure}

 We prepare a displaced triangular lattice consisting of $N\_{atom}=460$ atoms within an area of $L\times L \simeq 40a\times 40a$, where $a=1$ is the length unit
[top panel of Fig.~\ref{fig:Fourier}(a)].
 The atomic positions 
$\{X^\mathrm{true}_i,Y^\mathrm{true}_i\}$ 
($i=1,...N\_{atom}$) are treated as the hidden parameters $\bm{p}^\mathrm{true}$ to be predicted ($n=920$).
 An $M\times M$ real-space image $\{r^\mathrm{true}_{x,y}\}$ is generated by blurring each atomic position with a Gaussian of width $\sigma=0.3a$,
%___ eqn:r ___%
\begin{align}
 r^\mathrm{true}_{x,y} &= \sum_{i=1}^{N\_{atom}} \exp \! \left[ -\frac{(x-X^\mathrm{true}_i)^2 + (y-Y^\mathrm{true}_i)^2}{2\sigma^2} \right].
 \label{eqn:r}
\end{align}
%---%
%
 The corresponding $M\times M$ momentum-space image 
 $\{q^\mathrm{true}_{k_x,k_y}\}$
is obtained via Fourier transformation of $\{r^\mathrm{true}_{x,y}\}$.
 We set the image size $M = 128$
 ($N\_{data} = 16384$).
 For compatibility with AD, we use the \texttt{fft2} function from the JAX library, $\bm{q} = |\mathtt{jax.numpy.fft.fft2}(\bm{r})|/(M/2)^2$.
 For numerical stability, we define a quantity $\bm{s}$,
%___ eqn:s ___%
\begin{align}
 s_{k_x,k_y} &= L \ln \( q_{k_x,k_y} + 1\),
 \label{eqn:s}
\end{align}
%---%
for the target data and use the mean squared error [Eq.~\eqn{MSE}] between $\bm{s}$ and $\bm{s}^\mathrm{true}$ as the cost function.
 This quantity suppresses the influence of extreme values in the momentum-space image.
 In general, careful cost function engineering can significantly improve performance.

 As initial guesses, we use slightly displaced triangular lattices with the same number of atoms, assuming the density is known
 [top panel of Fig.~\ref{fig:Fourier}(b)].
 The momentum-space image $\{s_{k_x,k_y}\}$ is generated in the same way as for the target.
 For simplicity, we assume that Gaussian width $\sigma$ is known,
though it could also be treated as a hidden parameter.

 Figure~\ref{fig:Fourier}(c) shows an example of a predicted atomic configuration and its momentum-space image.
 The resulting lattice captures structural features of the true lattice,
and its Fourier image closely matches the target.
 If additional data or constraints---such as structure factors or excluded atomic radii---are available, they can be incorporated into the cost function to further refine the results.
 Despite its simplicity, the backsolution method has the potential to replace existing reverse modeling techniques, such as the reverse Monte Carlo method \cite{McGreevy01reverse}.

%===== FIG =====%
%===== FIG =====%

%%%%%%%%%%%%%%%%%%%
%%%%%  CONCL  %%%%%
%%%%%%%%%%%%%%%%%%%
\iftitle
 \section{Conclusion} \label{sec:conclusion}
 In this paper,
\else
 \textit{Conclusion.}
 In this Letter,
\fi
we have introduced a general-purpose framework for solving inverse problems using AD.
 The method, backsolution, solves inverse problems as parameter optimization tasks, analogous to backpropagation in neural networks,
 and enables efficient and scalable inference of hidden parameters from observational data.

 We demonstrated the effectiveness of backsolution by reconstructing spatial profiles (potential and magnetization landscapes) from magnetotransport data.
 This paves the way for the extraction of microscopic information from macroscopic transport data, potentially reducing the reliance on direct imaging techniques.
 Furthermore, we applied the method to generate disordered lattice structures whose Fourier-transformed images match a given target.
 This highlights the method's utility in reverse modeling, even when the available data are insufficient to uniquely determine the underlying configuration.

 While AD has previously been applied to specific optimization problems \cite{Yu21learning, Inui23inverse, Williams23automatic, Inui24inverse, Hirasaki24inverse},
its broader potential for solving inverse problems with many unknown parameters---such as reconstructing detailed sample profiles from experimental data---has been largely overlooked.
 One possible reason is the intuitive difficulty of inferring numerous non-smooth parameters from complicated data.
 Counterintuitively, as we have shown, the backsolution method remains effective as long as the cost function is smooth in the vicinity of the true solution.
 Importantly, our method does not require strong constraints, problem-specific hyperparameter tuning, or large training datasets, which often limit the applicability of existing approaches.
 Although we have kept the implementation intentionally simple, standard techniques from gradient-based optimization can be incorporated to further enhance performance.
 This highly versatile backsolution framework promises to solve a wide range of inverse problems.

\acknowledgments
 We thank R.~A.~R\"{o}mer for valuable discussions.
 This work was supported by JSPS
KAKENHI (Grant Nos.~%
JP22H05114 %TO Gakuhen-A
and
JP22K03446 %KK kiban-C
).

\bibliography{BackSolution}

\end{document}